\documentclass[aip,reprint,10pt,superscriptaddress,showbibnotes,showpacs]{revtex4-1}
\usepackage{amsfonts}
\usepackage{amsmath} 
\usepackage{amssymb} 
\usepackage{graphicx}
\usepackage{color}
\usepackage{subfigure}
\begin{document}
\title{Modulation of drift-wave envelopes in a nonuniform quantum magnetoplasma}
\author{A. P. Misra}
\email{apmisra@visva-bharati.ac.in, apmisra@gmail.com}
\affiliation{Department of Mathematics, Siksha Bhavana, Visva-Bharati University,  Santiniketan-731 235, West Bengal, India}
\begin{abstract}
We study the  amplitude modulation of  low-frequency, long-wavelength electrostatic drift-wave envelopes in a nonuniform quantum magnetoplasma consisting of cold ions and degenerate electrons. The effects of    tunneling associated with the quantum Bohm potential and the Fermi pressure for nonrelativistic degenerate electrons, as well as the  equilibrium density  and magnetic field inhomogeneities are taken into account. Starting from a set of quantum magnetohydrodynamic (QMHD) equations, we derive a    nonlinear Schr\"odinger equation (NLSE) that governs the dynamics of the modulated quantum drift-wave packets. The NLSE is used to study the modulational instability (MI) of a Stoke's wave train to  a small plane wave perturbation. It is shown that   the quantum tunneling effect as well as the scale length of inhomogeneity  play crucial roles for the MI of the  drift-wave packets.   Thus, the latter can propagate in the form of bright and dark envelope solitons or as  drift-wave rogons in degenerate dense magnetoplasmas. 
\end{abstract}
\maketitle
\section{Introduction} Degenerate dense plasmas have become a subject of important research over the last few years as those can be achieved in the laboratories \cite{dense-lab1,dense-lab2}, and can be useful for understanding the salient features of collective plasma oscillations in superdense astrophysical bodies like white dwarfs, neutron stars, magnetars etc. \cite{quantum-review}. {In these plasma environments, the typical number density of charged  particles (mainly electrons/positrons/holes) becomes extremely high, i.e., $n\gtrsim10^{26}$ cm$^{-3}$, and  their thermodynamic temperature  becomes low ($T\sim10^{5}-10^{7}$ K), so that the particles follow the Fermi-Dirac statistics. The degeneracy condition in these plasmas (where the Fermi energy is higher than the thermal energy) is satisfied for  $T\lesssim10^7$ K. However, in metals ($n\sim10^{23}$ cm$^{-3}$) electrons are degenerate at $T\lesssim10^5$ K.  Furthermore, in regimes, e.g., in an outer mantle of white dwarfs,  where electrons may be  nonrelativistic degenerate, the number density and the temperature satisfy $n\lesssim10^{26}$ cm$^{-3}$ and   $T\lesssim10^{7}$ K \cite{misra2012}, and the magnetic field is $B_0\lesssim10^8$ T. However, for ultra-relativistically degenerate electrons (e.g., in the core of white dwarfs) we can have the density $n\gtrsim10^{35}$ cm$^{-3}$ and the temperature   $T\lesssim10^{8}$ K \cite{misra2012}.   In degenerate dense plasmas there appear a time scale in the units of plasma period and a typical length scale $\lambda_F=v_F/\omega_p$, where $v_F$ is the Fermi velocity and $\omega_p$ is the plasma oscillation frequency. Just like the Debye length in classical plasmas, $\lambda_F$ represents the scale length of electrostatic screening in quantum plasmas. } In the latter,  the typical quantum mechanical  (Such as tunneling associated with the quantum Bohm potential) as well as the statistical (Fermi-Dirac pressure for degenerate species) effects play crucial roles for the collective plasma dynamics. Furthermore, in these high-density plasmas, the equilibrium  density of charged particles and the static ambient magnetic field  can be nonuniform with finite   scale lengths. Thus, there appear   the $\mathbf{E}\times\mathbf{B}$ drift (where ${\bf E}$ is the electric field and ${\bf B}$ is the magnetic field), diamagnetic drift and  polarization drift velocities due to the density and magnetic field inhomogeneities, as well as the  quantum drift  velocity caused by the strong density correlations due to quantum fluctuations.  It naturally becomes a challenge to understand the properties of drift waves as well as the formation of localized structures in degenerate dense plasmas \cite{drift-quantum,drift-quantum1,drift-quantum2,drift-quantum3,drift-quantum4,drift-quantum5,drift-quantum6}, as the quantum effects begin to modify the physical process at high densities.

Drift waves are typically low-frequency (compared to the ion gyrofrequency) electrostatic or electromagnetic  waves that  exist in a spatially nonuniform magnetoplasma. Such waves are caused by the guiding center drifts of the charged particles in presence of a density gradient across the external magnetic field. Various types  of drift waves, e.g.,  the electrostatic drift waves   and coupled drift-Alfv{\'e}n waves have been reported in classical [See, e.g., Refs. \cite{drift-class1,drift-class2}] as well as  quantum plasmas [See, e.g., Refs. \cite{drift-quantum,drift-quantum1,drift-quantum2,drift-quantum3,drift-quantum4,drift-quantum5,drift-quantum6}]  which play a crucial role in cross-field plasma particle transports \cite{r3,r4}, and the formation of coherent structures like solitons in space \cite{r5}, laboratory \cite{r6,r6a} as well as degenerate dense  plasmas that are magnetized.  Note that both the drift-Alfv\'en waves can be excited by free energy sources that are stored in the equilibrium pressure gradient and in magnetic field inhomogeneity. In astrophysical environments, since the parametric cascading of electromagnetic signals occurs, the drift waves may generate sufficient electromagnetic disturbance  that should be considered in telescopic observations. {Furthermore, since the cores or outer mantles of compact astrophysical  stars (e.g., white dwarfs) are  dense and  may consist  of a plasma of unbounded nuclei and electrons, i.e., a plasma composed of positively charged ions providing almost all the mass (inertia) and the pressure, as well as electrons providing the pressure (restoring force) but none of the mass (inertialess), and the magnetic field in these environments can be strong, the drift-wave excitation in degenerate dense plasmas could be useful for identifying the modulated drift-wave packets that may spontaneously emerge in magnetized white dwarfs, neutron stars etc.}
 { Tasso \cite{r7} and later Orevskii {\it et al.} \cite{r8} had investigated the nonlinear interactions of one-dimensional drift waves in presence of a uniform magnetic field as well as the density and temperature gradients in electron-ion plasmas. They reported the formation of  drift-wave solitons (non-envelope) that were used for the onset of drift-wave turbulence  \cite{r3,r9} in magnetized plasmas. However, two- or three-dimensional effects are essential for the drift-wave turbulence in classical, magnetically confined plasmas (tokamak) for the anomalous transport of charged particles.  } 
 
The modulational instability (MI) is   a  well-known mechanism for the energy localization of    wave packets in a  nonlinear dispersive  medium.  It  signifies the exponential growth of a small plane wave perturbation while propagating in  the medium. This gain leads to the amplification of the sidebands, and thereby breaking up the uniform wave into a train of pulses. Thus, the MI   acts as a precursor for the formation of bright envelope solitons or highly energetic rogue waves in plasmas \cite{drift-rogon,rogue-plasma,rogue-drift-plasma,rogue-alfven-plasma,rogue-peregrine}.  The amplitude modulation of a finite amplitude drift wave by zonal flows has been considered by Jovanovic {\it et al.} in a nonuniform magnetoplasma  \cite{modulation-zonal}. They showed that the full nonlinear system of equations governing drift-wave zonal flow interactions can be reduced to a cubic nonlinear Schr\"odinger equation (NLSE), which possesses localized envelope soliton solutions. {It is to be noted that the nonlinear interaction between zonal flows and drift waves as in Ref. \cite{modulation-zonal}   involves  one-dimensional propagation that belong to the convective cell mode (with a non-Boltzmannean electron distribution) rather than to the drift-mode spectrum. Though, such nonlinear interaction is energetically more favorable than one with zonal flows that belong to the drift-wave spectrum, there may be the situation when the convective-cell part of the spectrum does not appear, e.g., in the dynamics of Rossby-wave turbulence in rotating fluids \cite{jovanovic2010}. However, it has been shown that the presence of  immobile charged dust grains may lead to the appearance of an additional term (proportional to the dust-density gradient) in the ion vorticity equation [See, e.g., Eq. (13) in Ref. \cite{drift-quantum}] which is associated with the Shukla-Varma mode (convective cell) \cite{shukla-varma} in degenerate dense electron-ion plasmas. Furthermore, the convective cell mode is not affected by the degeneracy pressure of electrons and the quantum force associated with the Bohm potential.}  {So, in  degenerate electron-ion plasmas with no charged dust, the quantum effects    may appear only in the drift-wave part of the spectrum.   } 

Recently, the nonlinear properties of modulated one-dimensional drift-wave packets in a nonuniform magnetoplasma with the effects of equilibrium density, electron temperature and magnetic field gradients have been considered by Shukla {\it et al} \cite{drift-rogon}. It was shown that the dynamics of the modulated drift-wave packet is governed by a NLSE, which depicts the formation of dark  and bright solitons, as well as drift-rogue waves.   Furthermore, the nonlinear theory of  cylindrical lower-hybrid drift-solitary waves in an inhomogeneous, magnetized plasma with the   effects of electron density  and   temperature inhomogeneities has been studied by Liu {\it et al.} \cite{liu-drift} in a two-fluid model. They reported a diminution of the wave amplitude and width of the solitary waves with the enhancement of  the density inhomogeneity.

In this paper, we study the  amplitude modulation of one-dimensional quantum drift-wave packets in a nonuniform magnetoplasma with the effects of equilibrium density  and magnetic field gradients as well as the quantum force associated with the Bohm potential and the degeneracy pressure of electrons. It is shown that the dynamics of the modulated drift-wave packet is governed by a modified NLSE, which depicts the formation of dark and bright envelope solitons, as well as drift-wave rogons.         
 
\section{Derivation of quantum drift-wave equation} We consider a nonuniform quantum electron-ion magnetoplasma in  presence of the inhomogeneities  of the equilibrium density and the  external magnetic field $\hat {\bf z} B_0 (x)$. Thus, at equilibrium, the gradient of the total energy (Fermi energy density plus the magnetic energy density) vanishes, i.e., 
\begin{equation}
\frac{\partial}{\partial x}\left[n_0 (x) T_{Fe} (x) + \frac{B_0^2(x)}{8\pi}\right]=0,
\label{eqbm}
\end{equation}
where $n_0 (x)$ is the unperturbed electron or ion number density and $T_{Fe} (x)\equiv(\hbar^2/2k_Bm_e)\left(3\pi^2n_0(x)\right)^{2/3}$  is the electron Fermi temperature in which $\hbar=h/2\pi$ is the reduced Planck's constant, $m_e$ is the electron mass and $k_B$ is the Boltzmann constant.

In the propagation of the low-frequency (in comparison with the ion gyrofrequency $\omega_{ci} =eB_0/m_i c$, where $e$ is the magnitude of the electron charge, $m_i$ is the ion mass, and $c$ is the speed of light in vacuum),
long-wavelength (in comparison with the ion-thermal gyroradius $\rho_i =c_{s}/\omega_{ci}$, where  $c_{s} =\sqrt{2k_B T_{Fe}/m_i}$ is the quantum ion-acoustic speed) 
electrostatic (with the field ${\bf E}=-\nabla \phi$, where $\phi$ is the electrostatic potential)    drift waves in nonuniform quantum magnetoplasmas, the perpendicular 
(to $\hat {\bf z}$) components of the electron and ion fluid velocities are \cite{drift-quantum,drift-rogon}

\begin{eqnarray}
{\bf v}_{e\perp}\approx&&\frac{c}{B_0(x)}\hat {\bf z} \times \nabla_{\perp} \phi -\frac{c} 
{eB_0(x)n_e}\hat {\bf z} \times \nabla P_e \notag\\
&&-\frac{c}{eB_0(x)}\hat {\bf z}\times\nabla \Psi_q\equiv\mathbf{V}_E+\mathbf{V}_D+\mathbf{V}_Q, \label{ev}\\
{\bf v}_{i\perp} \approx&& \frac{c}{B_0(x)}\hat {\bf z} \times \nabla_{\perp} \phi-\frac{c}{B_0 (x) \omega_{ci}}\frac{d}{dt} \left(\nabla_\perp \phi\right)\notag\\ &&\equiv\mathbf{V}_E+\mathbf{V}_P, \label{iv}
\end{eqnarray}
where we have assumed that $|d/dt| \ll \nu_{ei} \ll \omega_{ce}$, with $ d/dt \equiv \partial/\partial t + {\bf V}_E \cdot \nabla$ and $\nabla_{\perp}\equiv\hat{x}(\partial/\partial x)+\hat{y}(\partial/\partial y)$. Here $\nu_{ei}$ is the electron-ion collision frequency and $ \omega_{ce} =eB_0/m_e c$ is the electron gyrofrequency. Also, $\Psi_q=-\left(\hbar^2/2m_e\right)\left(\nabla^2\sqrt{n_e}/\sqrt{n_e}\right)$ is the quantum Bohm potential, and  $\mathbf{V}_E$, $\mathbf{V}_D$, $\mathbf{V}_P$ and $\mathbf{V}_Q$ are, respectively,  the $\mathbf{E}\times\mathbf{B}$ drift, diamagnetic drift, polarization drift and the  quantum drift (due to tunneling) velocities.   The pressure $P_e$ for weakly relativistic degenerate electrons is given by the following equation of state \cite{degenerate-pressure1,degenerate-pressure2}
\begin{equation}
P_e=\frac{m_eV^2_{Fe}}{5n^{2/3}_0(x)}n^{5/3}_e\equiv\frac{1}{5}\left(3\pi^2\right)^{2/3}\frac{\hbar^2}{m_e}n_e^{5/3},\label{pressure-equation}
\end{equation}    
where $V_{Fe}=\sqrt{2k_BT_{Fe}/m_e}$  is the electron Fermi-thermal speed. 

Next, we neglect the ion motion parallel to $\hat{\bf z}$ as well as the compressional magnetic field perturbation  (i.e., we discard the coupling between the drift waves and the quantum ion-acoustic waves) \cite{drift-quantum,drift-rogon}. {Such an approximation is  valid in a low-$\beta$ plasma with wave frequency satisfying $|\partial/\partial t|\gg v_F|\partial/\partial z|$.  } Thus, inserting the parallel component of the electron fluid velocity, i.e., $v_{ez}\approx(c/4\pi en_e)\nabla^2_{\perp}A_z$, where $A_z$ is the parallel (to $\hat{\bf z}$) component of the vector potential such that ${\bf B}_{\perp}=\nabla A_z\times\hat{\bf z}$, into the parallel component of the inertialess electron momemntum equation, we obtain for $dA_z/dt,\lambda_e^2(\partial/\partial t)\nabla^2_{\perp}A_z\ll\partial\phi/\partial z$, where $\lambda_e=c/\omega_{pe}$ is the collisionless electron skin depth with $\omega_{pe}$ denoting the electron plasma oscillation frequency,  
the following modified (by the quantum effects) Boltzmann law for the electron number density perturbation \cite{boltzmann-quantum}
\begin{eqnarray}
\frac{n_{e1}}{n_{0}(x)}=&&\left[1+\varphi+\frac{\mathcal{H}^2}{2}\left(1+\varphi\right)^{-3/4}\nabla^2 \left(1+\varphi\right)^{3/4}\right]^{3/2}\notag\\
&&\approx1 +\frac{3}{2}\varphi +\frac{3}{8}\varphi^2\notag\\
&&+\frac{9\mathcal{H}^2}{16}\left[\nabla^2\varphi-\frac{1}{4}(\nabla\varphi)^2-\frac{1}{2}\varphi\nabla^2\varphi\right],\label{electron-eqn}
\end{eqnarray}  
where  $n_{e1}\ll n_0$, $\varphi=e\phi/k_BT_{Fe}\ll1$ and $\mathcal{H}=H\omega_{ci}/\omega_{pi}$ with $\omega_{pi} =(4\pi n_0e^2/m_i)^{1/2}$ denoting the ion plasma oscillation frequency and  $H=\hbar\omega_{pe}/k_BT_{Fe}$  the ratio of the plasmon energy density to the Femi energy density. In deriving Eq. \eqref{electron-eqn}, we have considered the semiclassical limit $\mathcal{H}^2\ll1$.

Substituting Eq. \eqref{iv} into the ion continuity equation, we obtain \cite{drift-rogon}
\begin{equation}
\frac{\partial n_{i1}}{\partial t} + \nabla \cdot [(n_0 (x) + n_{i1}) {\bf V}_E ] 
\approx \frac{c n_0(x)}{B_0(x) \omega_{ci}} \frac{d \nabla_\perp^2 \phi}{dt}.
\label{ion-eqn}  
\end{equation}
We can  now combine Eqs. \eqref{electron-eqn} and \eqref{ion-eqn}  under the quasi-neutrality condition $n_{i1} = n_{e1}$, which holds for a dense magnetized plasma
with $\omega_{pi} \gg \omega_{ci}$ to obtain the  modified quantum drift-wave equation in one-space dimension 

\begin{eqnarray}
&& \left(1-\frac{\partial^2}{\partial y^2}\right)\frac{\partial \varphi}{\partial t}+\left(\alpha-\beta \mathcal{H}^2\frac{\partial^2\varphi}{\partial y^2}\right) \frac{\partial \varphi}{\partial y}
+ \frac{1}{2}\varphi \frac{\partial \varphi}{\partial t}-\gamma \varphi \frac{\partial \varphi}{\partial y}\notag\\
&&+\delta \mathcal{H}^2\frac{\partial}{\partial t}\left[\left(1-\frac12\varphi\right)\frac{\partial^2 \varphi}{\partial y^2}-\frac14\left(\frac{\partial \varphi}{\partial y}\right)^2\right]=0, 
\label{driftwave}
\end{eqnarray} 
where 
\begin{equation}
\alpha=-\sqrt{\frac{2}{3}}\frac{\partial}{\partial x}\ln\left[\frac{n_0(x)}{B_0(x)}\right]>0,
\end{equation}
\begin{equation}
\beta=\frac{1}{4}\left(\frac{3}{2}\right)^{5/2}\frac{\partial}{\partial x}\ln\left[\frac{n^{-1/3}_0(x)}{B_0(x)}\right],
\end{equation}
\begin{equation}
\gamma=\sqrt{\frac{3}{2}}\frac{\partial}{\partial x}\ln\left[\frac{n^{1/3}_0(x)}{B_0(x)}\right],
\end{equation}
and $\delta=9/16$. In Eq. \eqref{driftwave}, the time $t$ is normalized by the ion gyroperiod $\omega^{-1}_{ci}$, the space coordinates $x$    and $y$ are normalized by $\rho_s$ and  $\sqrt{(2/3)}\rho_s$ respectively.  Furthermore, the drift-wave equation \eqref{driftwave} has been modified due to both the quantum mechanical and statistical effects. The latter also modify the expressions for $\alpha,~\beta$ and $\gamma$, which appear due to the density and magnetic field gradients, and can change their signs with the choice of the scale lengths of inhomogeneity.     In the formal limit of $\mathcal{H}=0$, i.e., simply disregarding the quantum Bohm potential term (tunneling effect), one can recover the similar expression as in Ref. \cite{drift-rogon} except  some factors which appear due to the different pressure law for degenerate electrons.

\section{Derivation of NLSE and modulational instability} We derive the governing nonlinear equation for the  amplitude-modulated low-frequency $(\omega<\omega_{ci}\ll\omega_{pi})$ drift-wave packets in quantum magnetoplasmas. {Note that the wave equation \eqref{driftwave}, which describes the evolution of the electrostatic perturbation $\varphi$ has a harmonic wave solution   $\varphi=\varphi_0\exp(iky-i\omega t)$  in the small-amplitude limit $\varphi_0\ll1$. However, whenever the wave amplitude becomes non-negligible, a nonlinear mechanism for the generation of harmonic waves comes into play. In order to study the amplitude modulation and associated stability/ instability profiles of these electrostatic drift waves we assume that $\varphi$ takes the form of a modulated wave packet, i.e., the composition  of a fast carrier wave with a slow variation in amplitude. Initial drift-wave packets are modulated by the nonlinear effects. If we see the packet on the coordinate frame   moving at the group velocity $v_g$ (to be determined by the linear dispersion relation), the time variation of the packet looks slow and hence the space and the time variables are stretched as $\xi=\epsilon (y-v_gt)$, $\tau=\epsilon^2 t$, where $\epsilon$ is a small free (real) parameter ($0 < \epsilon \ll 1)$ representing the weakness of perturbation. 
Following the standard reductive perturbation technique \cite{rpt}, we expand $\varphi$ about its equilibrium value as}
{
\begin{equation}
\varphi=0+\epsilon\varphi^{(1)}+\epsilon^2\varphi^{(2)}+\cdots,
\end{equation}
where the perturbation $\varphi^{(n)},~n=1,2,3,...$ can be considered as a sum of infinite number of Fourier modes:}
{
\begin{equation} 
\varphi^{(n)}=\sum^{\infty}_{l=-\infty}\varphi_l^{(n)}(\xi,\tau)\exp[il(k y-\omega t)].
\end{equation} }
{
Thus, we express the wave potential $\varphi$ as
\begin{equation} 
\varphi=\sum^{\infty}_{n=1}\epsilon^n \sum^{\infty}_{l=-\infty}\varphi_l^{(n)}(\xi,\tau)\exp[il(k y-\omega t)], 
\label{expansion}
\end{equation} 
where $\varphi_{-l}^{(n)}=\varphi_{l}^{(n)*}$ holds for real physical variables and the asterisk denotes the complex conjugate. In practice, only the terms with $l\leq n$ contribute in the summation, i.e.,  (up to) first harmonics are expected for $n=1$, up to second harmonics for $n=2$ etc. }   

 {Next,  to derive an evolution equation of  NLSE-type, we follow   the similar method as, e.g., in Refs. \cite{kourakis2005,Mckerr2014}. In Ref.  \cite{kourakis2005}, the modulational instability and the evolution of localized wave envelopes in space and dusty plasmas have been studied. Here, starting from a set of fluid equations, and using a standard reductive perturbation technique,   a NLSE has been derived which was shown to admit bright or dark envelope solitons. Furthermore, the occurrence of freak waves or rogons associated with the   propagation of electrostatic wave packets in quantum electron-positron-ion plasmas has   been investigated by Mckerr {\it et al.} \cite{Mckerr2014}. Using a multiscale technique, the authors have shown that the evolution of the wave envelopes can be described by a NLSE, which admits envelope solitons as well as localized breathers. Thus, following Refs. \cite{kourakis2005,Mckerr2014} we substitute the expansion Eq. \eqref{expansion} into Eq. \eqref{driftwave}, and equate different powers of $\epsilon$. }

  For $n=l=1$, equating the coefficient of $\epsilon$, we obtain the following  linear dispersion law for the quantum drift waves {(Since we are interested in the modulation of a plane wave with frequency $\omega$ and wave number $k$, we put $\varphi_{l}^{(1)}=0$ for all $l$ except $l=\pm1$)} 
\begin{equation}
\omega =\frac{\alpha k}{1+k^2\left(1-\delta\mathcal{H}^2\right)}.
\label{DR}
\end{equation}

Since $\delta\mathcal{H}^2<1$, it follows that the frequency of the carrier drift-wave mode decreases as the quantum parameter $\mathcal{H}~(<1)$ decreases from a certain value, say $\mathcal{H}=0.3$. This is the consequence of relatively high-density regimes for plasmas in which a drift-wave carrier modes  propagate  with  lower frequencies. The effect of $\mathcal{H}$ on the wave modes is more pronounced when the wave numbers $k$   approaches $1$. However, in the limit $k\rightarrow0$,   the wave becomes dispersionless.
   
From the second-order expressions for the first harmonics, i.e., for $n=2,l=1$, we obtain an equation in which the coefficient of $\varphi_1^{(2)}$ vanishes by the dispersion equation \eqref{DR}, 
and the coefficient of $\partial \varphi_1^{(1)}/\partial\xi$, after equating to zero, gives the following compatibility condition:

\begin{equation} 
v_g\equiv\frac{\partial\omega}{\partial k}=\frac{\alpha\left(1-k^2+\delta\mathcal{H}^2k^2\right)}{\left(1+k^2-\delta\mathcal{H}^2k^2\right)^2}. \label{vg}
\end{equation} 
Thus, the group velocity of the drift-wave packet is also modified by the term $\propto~\mathcal{H}$. Its value increases (decreases) with increasing (decreasing) values of $\mathcal{H}$. Evidently, $v_g>0$ for $\alpha>0,~\delta\mathcal{H}^2<1$ and for wave numbers satisfying $k<1/\sqrt{1-\delta\mathcal{H}^2}$.  

The zeroth harmonic mode   appears  due to the nonlinear self-interaction of the drift-wave modes. Thus, equating   the coefficient  of $\epsilon^3$ for $n=2,l=0$, we obtain  
\begin{equation}
\varphi_0^{(2)}=\left(\frac{\gamma+v_g/2+\beta\mathcal{H}^2k^2}{\alpha-v_g}\right)\left\vert\varphi_1^{(1)}\right\vert^2\equiv \lambda\left\vert\varphi_1^{(1)}\right\vert^2. \label{zerothorder}
\end{equation}

Next, we consider the second-order harmonic mode for $n=l=2$,  and equate the coefficient of $\epsilon^2$ to obtain

\begin{eqnarray}
\varphi_2^{(2)}=&&\frac{\omega+2\gamma k+2\mathcal{H}^2k^2\left(3\omega\delta/2-\beta k\right)}{4\left[ \alpha k-\omega(1+4k^2)+2\omega\delta\mathcal{H}^2k^2\right]}\left[\varphi_1^{(1)}\right]^2\notag\\
&&\equiv \mu\left[\varphi_1^{(1)}\right]^2.\label{2ndorder}
\end{eqnarray}

Finally, for $n=3, l=1$, we obtain an equation for the third-order first-harmonic mode  in which the coefficients of $\varphi^{(3)}_1$ 
and $\partial \varphi_1^{(2)}/\partial\xi$ vanish by the dispersion relation and the group velocity expression, respectively. 
In the reduced equation we substitute the expressions for $\varphi_0^{(2)}$ and $\varphi_2^{(2)}$ from Eqs. \eqref{zerothorder} and \eqref{2ndorder} to obtain the 
following NLSE  
\begin{equation}
i\frac{\partial \Phi}{\partial \tau}+P\frac{\partial^2 \Phi}{\partial \xi^2}+Q|\Phi|^2\Phi=0, \label{NLSE}
\end{equation}
where $\Phi=\varphi^{(1)}_1$ is the potential perturbation, or in the original frame of reference

\begin{equation}
i\left(\frac{\partial }{\partial t}+v_g\frac{\partial}{\partial y} \right)\Phi 
+ P\frac{\partial^2 \Phi}{\partial y^2}+Q|\Phi|^2\Phi=0,
\end{equation}
with $\Phi\sim \epsilon \varphi^{(1)}_1$. The coefficients of the drift-wave group dispersion and the nonlinearity are

\begin{equation}
P \equiv\frac{1}{2}\frac{\partial^2 \omega}{\partial k^2}= \frac{\left(\delta\mathcal{H}^2-1\right)\left(\omega+2kv_g\right)}{\left(1+k^2-\delta\mathcal{H}^2k^2\right)},\label{P-dispersion}
\end{equation}
and
\begin{equation}
Q=\frac{(\lambda+\mu)\left(\omega/2+\gamma k\right)+\mathcal{H}^2k^2Q_0}{1+k^2-\delta\mathcal{H}^2k^2}, 
\label{Q-nonlinear}
\end{equation}
where $Q_0=2\beta\mu k+\omega\delta(\lambda+7\mu)/2$.
\begin{figure}[ht]
\centering
\includegraphics[height=2.0in,width=3.5in]{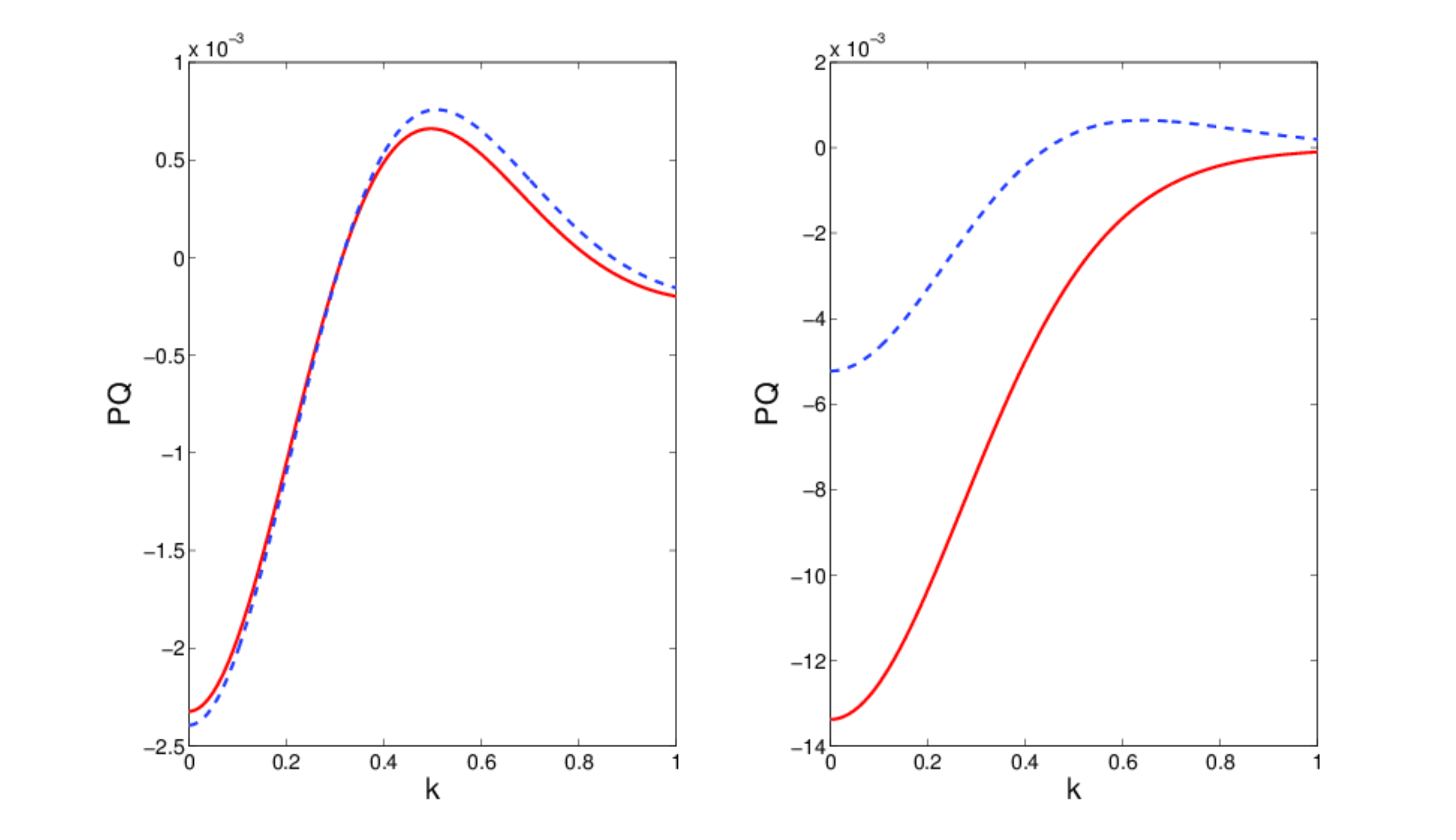}
\caption{The  stable $(PQ<0)$ and unstable $(PQ>0)$ domains with respect to $k$ are shown. The left panel shows the domains for different values of $\mathcal{H}$:  $\mathcal{H}=0.1$ (solid line) and  $\mathcal{H}=0.3$ (dashed line). The scale lengths of inhomogeneities are $L_n=2,~L_b=6L_n$ for which $\alpha=0.34,~\beta=0.17$ and $\gamma=-0.1$. The right panel shows the same, but  for different values of the scale length of inhomogeneity: $L_b=2.5L_n$  (solid line) for which $\alpha=0.24,~\beta=0.25$ and $\gamma=0.04$, and  $L_b=4L_n$ (dashed line) for which $\alpha=0.3,~\beta=0.2$ and $\gamma=-0.05$. The  other fixed parameters are  $L_n=2$  and $\mathcal{H}=0.1$. Thus, the quantum effects favor the instability, while the reduced scale length of inhomogeneity with $\gamma>0$ favors the stability of the drift-wave packets.}
\label{fig:fig-PQ}
\end{figure}  

In a generic manner, the modulated drift-wave packet whose amplitude is governed through the NLSE \eqref{NLSE} can be stable (unstable) to a plane wave perturbation if $PQ<0~(>0)$ (See, e.g., Ref. \cite{mi-review,drift-rogon}). From Eq. \eqref{P-dispersion} we find that 
$P$ is always negative for $\delta\mathcal{H}^2<1$. Thus, the sign of $PQ$ changes with only     the sign change of the nonlinear coefficient $Q$. The latter depends not only on the range of values of the wave number $k<1$, but also on the quantum parameter $\mathcal{H}$ as well as the  scale lengths of inhomogeneities. It turns out that the stable/ unstable domains of $k$ get modified with the inclusion of the quantum effects. Following, e.g., Ref. \cite{drift-rogon} we find that to a plane wave perturbation with frequency $\Omega$ and wave number $K$, the product $PQ\gtrless0$ according to when $K\lessgtr K_c$, where $K_c\equiv \sqrt{2|Q/P|}|\Phi_0|$ is the critical value of $K$ with $\Phi_0$ denoting the potential of the drift-wave pump. The instability growth  rate is given by
\begin{equation}
\Gamma= |P| K^2\sqrt{\frac{K^2_c}{K^2}-1}, \label{instability-rate}
\end{equation}
with a maximum  $\Gamma_{\text{max}}=|Q||\Phi_0|^2$.  

Next, we numerically investigate the range of values of $k$ in which the drift waves become stable or unstable to the modulation for different values of the parameter $\mathcal{H}$ as well as the scale length of inhomogeneity. For convenience, we assume the equilibrium  density and the magnetic field with the scale lengths $L_n$ and $L_b$  to be of the  forms $n_0(x)=n_0(0)\exp\left(-x/L_n\right)$  and $B_0(x)=B_0(0)\exp\left(-x/L_b\right)$ respectively. We also assume that $L_b>L_n$ so that $\alpha>0$. The sign of $\gamma$ will depend on the choice of values of $L_b$ larger than $L_n$.   
\begin{figure}[ht]
\centering
\includegraphics[height=2.0in,width=3.5in]{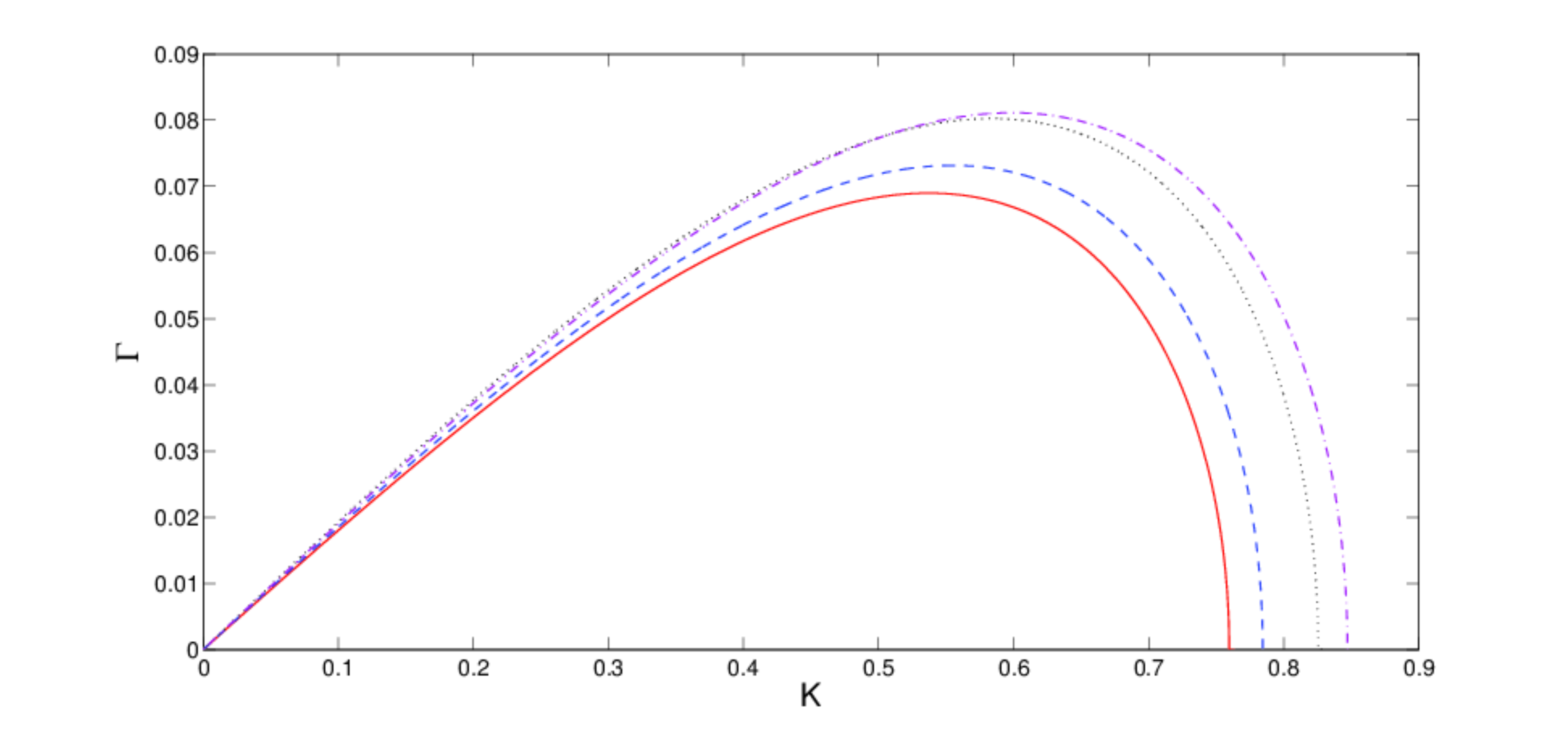}
\caption{The instability growth rate given by Eq. \eqref{instability-rate} is shown at $k=0.5$ for different values of the quantum parameter $\mathcal{H}$: $\mathcal{H}=0.1$ (solid line), $0.2$ (dashed) and $0.3$ (dotted), and for   fixed   $L_n=2,~L_b=6L_n$ and $\Phi_0=5$.  The  dash-dotted line corresponds to the case of reduced scale length of inhomogeneity (compared to the dotted line):  $L_n=2,~L_b=5L_n$ with $\mathcal{H}=0.3$ and $\Phi_0=5$.   }
\label{fig:fig-growth}
\end{figure}  
The stable $(PQ<0)$ and unstable $(PQ>0)$ regions are plotted against $k$ as shown in Fig. \ref{fig:fig-PQ}. It is seen that at higher values of $\mathcal{H}$, which correspond   to   relatively   low-density regimes, the instability  occurs in a larger domain of $k$. However, the stable regions remain almost unchanged with $\mathcal{H}$. We find that for $\mathcal{H}=0.1$, $PQ>0$ occurs in $0.3149\lesssim k\lesssim0.83$, while $0.3148\lesssim k\lesssim0.88$ represents  the domain in which $PQ>0$ for $\mathcal{H}=0.3$. Typically, for a  ratio $\omega_{ci}/\omega_{pi}=0.1$, the value $\mathcal{H}=0.1$ corresponds to the plasma number density   $n_0\sim10^{24}$ cm$^{-3}$, while $n_0\sim10^{21}$ cm$^{-3}$ is for $\mathcal{H}=0.3$.       On the other hand, the larger the scale length of inhomogeneity of the magnetic field, the greater is the instability domain. From Fig. \ref{fig:fig-PQ} it is also found that there exists a critical value of the scale length $L_b$ below which the drift wave is modulationally stable.  

 The MI growth rate $\Gamma$ is exhibited in Fig. \ref{fig:fig-growth}.  We find that the growth rate of instability is enhanced with higher values of $\mathcal{H}$, i.e., as one enters from higher density regimes (say, $n_0\sim10^{24}$ cm$^{-3}$) to relatively lower ones  (say, $n_0\sim10^{21}$ cm$^{-3}$), $\Gamma$ tends to attain its maximum value. In other words, the growth rate of instability is suppressed when the quantum effects are more pronounced in high-density regimes. The increasing trend of $\Gamma$ is also seen for a fixed $\mathcal{H}$, but at lower values of the scale length of inhomogeneity $L_b$. In each of these cases, the    cut-offs of $\Gamma$ occur   at higher values of the wave number of modulation $K$.  
 
In the ranges of values of $k$ for which the drift-wave packet becomes unstable $(PQ>0)$,  Eq. \eqref{NLSE} can be rewritten as
\begin{equation}
i\frac{\partial \Phi}{\partial \tilde{\tau}}+\frac12\frac{\partial^2 \Phi}{\partial \tilde{\xi}^2}+|\Phi|^2\Phi=0, \label{NLSE-nondim}
\end{equation}
where $\tilde{\tau}=Q\tau$ and $\tilde{\xi}=\sqrt{Q/2P}\xi$. 
Thus, Eq. \eqref{NLSE-nondim} have the following rogue wave solution  that is located on a non-zero background and localized  in  both  space and time \cite{2nd-order-rogue-coeff}  
\begin{equation}
\Phi_{n}(\tilde{\xi},\tilde{\tau})=\left[(-1)^n+\frac{G_n(\tilde{\xi},\tilde{\tau})+iH_n(\tilde{\xi},\tilde{\tau})}{D_n(\tilde{\xi},\tilde{\tau})} \right]\exp(it), \label{rogue-solution-general}
\end{equation}
where $G_n,~H_n$ and $D_n~(\neq0)$ are some polynomial functions of $\tilde{\xi}$ and $\tilde{\tau}$, and $n=1,2,3,\cdots$, is the order of the solution.  The first-order $(n=1)$ solution  corresponds to the Peregrine soliton \cite{peregrine-soliton} in which $G_1=4$, $H_1=8\tilde{\tau}$ and $D_1=1+4\tilde{\xi}^2+4\tilde{\tau}^2$. However,   superposition of two first-order rogue waves is also possible, and can cause the generation of another highly energetic rogue waves with higher amplitudes. The analytic form of these  waves have  been recently obtained by  Akhmediev \textit{et al.} \cite{rogue-2nd-order}  using the deformed Darboux transformation approach. This second order $(n=2)$ rogon solution has the same form as Eq. \eqref{rogue-solution-general}    where  the polynomials $G_2$, $H_2$ and $D_2$   are given by \cite{2nd-order-rogue-coeff,rogue-2nd-order}
\begin{equation}
G_2=-\left(\tilde{\xi}^2+\tilde{\tau}^2+\frac{3}{4}\right)\left(\tilde{\xi}^2+5\tilde{\tau}^2+\frac{3}{4}\right)+\frac{3}{4},  
\end{equation}
\begin{equation}
H_2=\tilde{\tau}\left[3\tilde{\xi}^2-\tilde{\tau}^2-2\left(\tilde{\xi}^2+\tilde{\tau}^2\right)^2-\frac{15}{8}\right],
\end{equation} 
\begin{equation}
D_2=\frac{1}{3}\left(\tilde{\xi}^2+\tilde{\tau}^2\right)^3+\frac{1}{4}\left(\tilde{\xi}^2-3\tilde{\tau}^2\right)^2+
\frac{3}{64}\left(12\tilde{\xi}^2+44\tilde{\tau}^2+1\right).
\end{equation} 
\begin{figure}[ht]
\centering
\subfigure[]{\includegraphics[height=2.0in,width=3.5in]{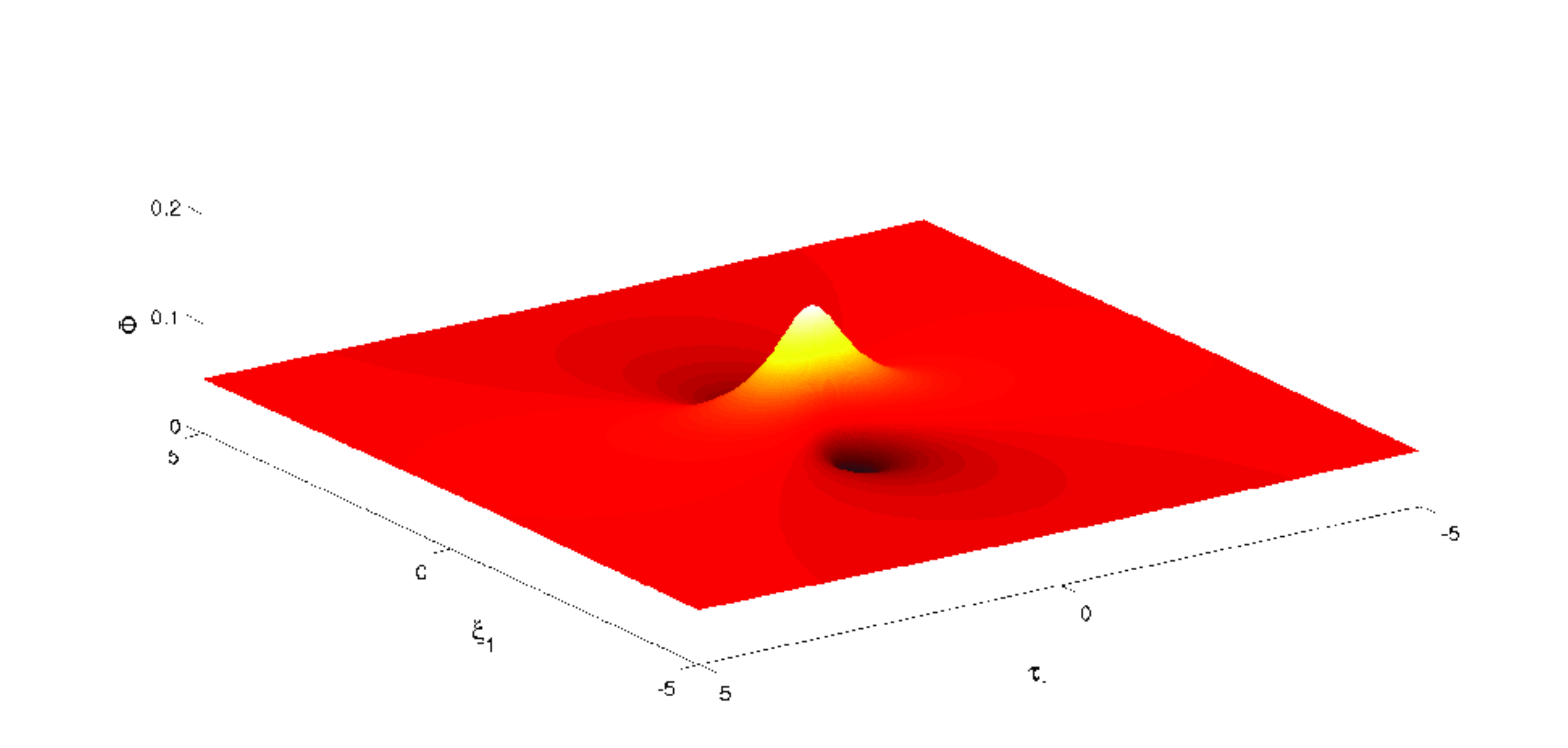}
\label{fig:subfig1}}
\subfigure[]{\includegraphics[height=2.0in,width=3.5in]{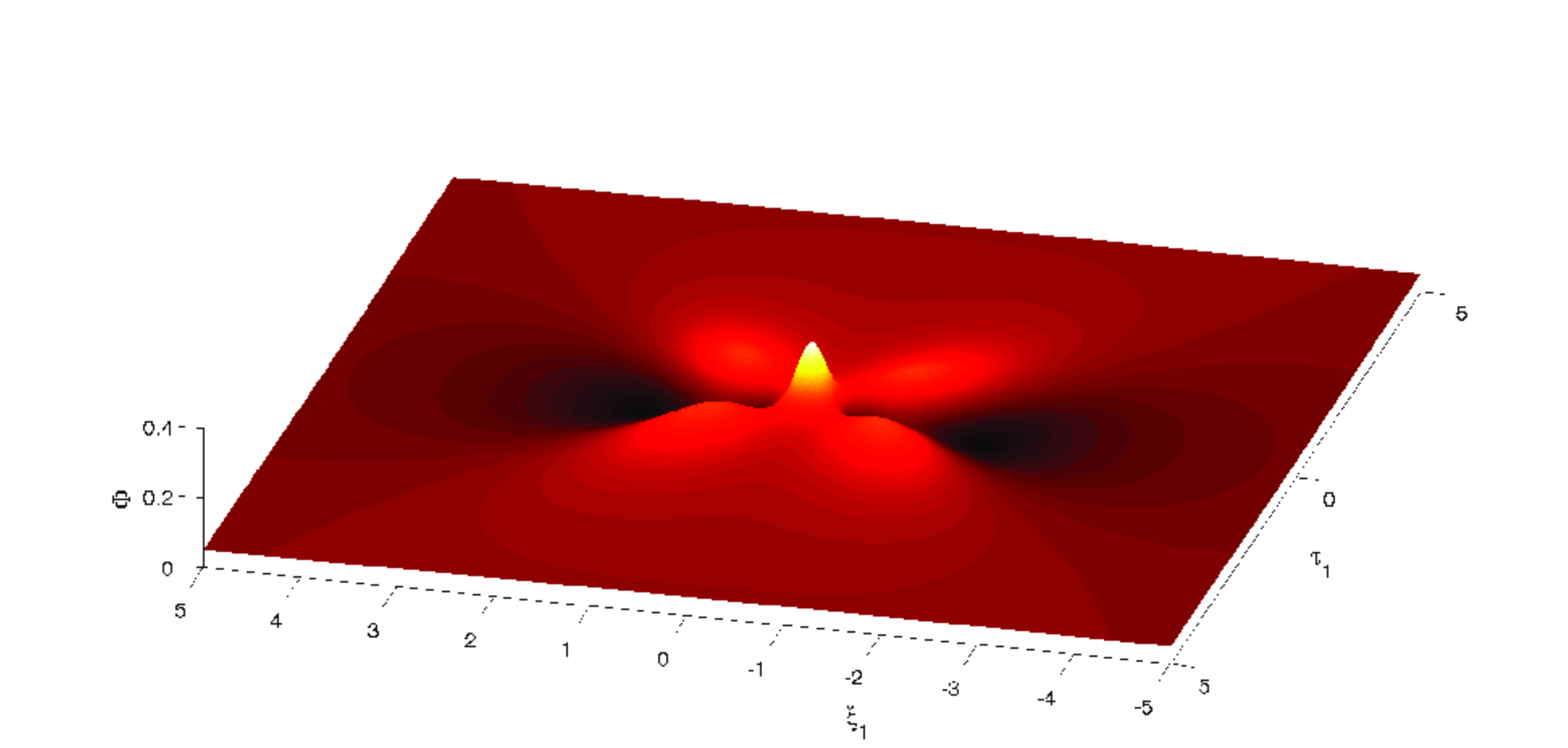}
\label{fig:subfig2}}
\caption{The evolution of (a) first (upper panel) and (b) second (lower panel) order rogon solutions of Eq. \eqref{NLSE-nondim} at $k=0.5$ for $L_n=0.01,~L_b=6L_n$ and   $\mathcal{H}=0.1$. Here, $\xi_1=\tilde{\xi}$ and $\tau_1=\tilde{\tau}$. The corresponding  values of $P$, $Q$, $\alpha,~\beta$ and $\gamma$ are  $P=-0.2391$, $Q=-0.0028$, $\alpha=68,~\beta=34$ and $\gamma=-20$. }
\label{fig:fig-rogon}
\end{figure}  
 The typical forms of these rogon solutions given by Eq. \eqref{rogue-solution-general} for $n=1,2$  are shown in Fig.  \ref{fig:fig-rogon}.   These highly energetic rogue waves in which a significant amount of energy is concentrated   in  a relatively small area in  space and time, are  generated due to the MI of the   coherent  quantum drift-wave packets in the limit of infinite wave modulation period. In particular, they significantly amplify the  carrier wave amplitudes, and hence increase the nonlinearity during the evolution of the  wave packets. The  amplification factor of the amplitude of the $n$-th order rational solution [Eq. \eqref{rogue-solution-general}]  at $\tilde{\xi}=\tilde{\tau}=0$ is, in general, of $2n + 1$. Hence, localized quantum drift waves that are modelled by the higher-order breather solutions can also cause the formation of super rogue waves. To mention, the first-order Peregrin soliton has been recently observed experimentally in   plasmas   \cite{rogue-peregrine}. However,   the second-order rogon solution, which has been observed in water waves \cite{2nd-order-rogue-water}, is yet to  observe in plasmas.

As is known, MI of wave envelopes also gives rise to the formation of bright envelope solitons of Eq. \eqref{NLSE}, whose exact analytic form can be obtained by considering $\Phi=\sqrt{\Psi}\exp(i\theta)$, where $\Psi$ and $\theta$ are real functions to be determined (see for details, e.g., Ref. \cite{soliton-solutions}). Thus, for $PQ>0$ Eq. \eqref{NLSE} has the following bright-envelope soliton solution 
\begin{equation}
\Psi=\Psi_{b0} \hskip1pt\text{sech}^2\left(\frac{\xi-U\tau}{W_b}\right),\hskip2pt \theta=\frac{1}{2P}\left[U\xi+\left(\Omega_0-\frac{U^2}{2}\right)\tau\right].\label{bright-envelope}
\end{equation} 
This represents a localized pulse traveling at a speed $U$ and oscillating at a frequency  $\Omega_0$ at rest. The   width $W_b$  of the pulse is  given by $W_b=\sqrt{2P/Q\Psi_{b0}}$, where $\Psi_{b0}$ is the constant amplitude.
\begin{figure}[ht]
\centering
\includegraphics[height=2.0in,width=3.5in]{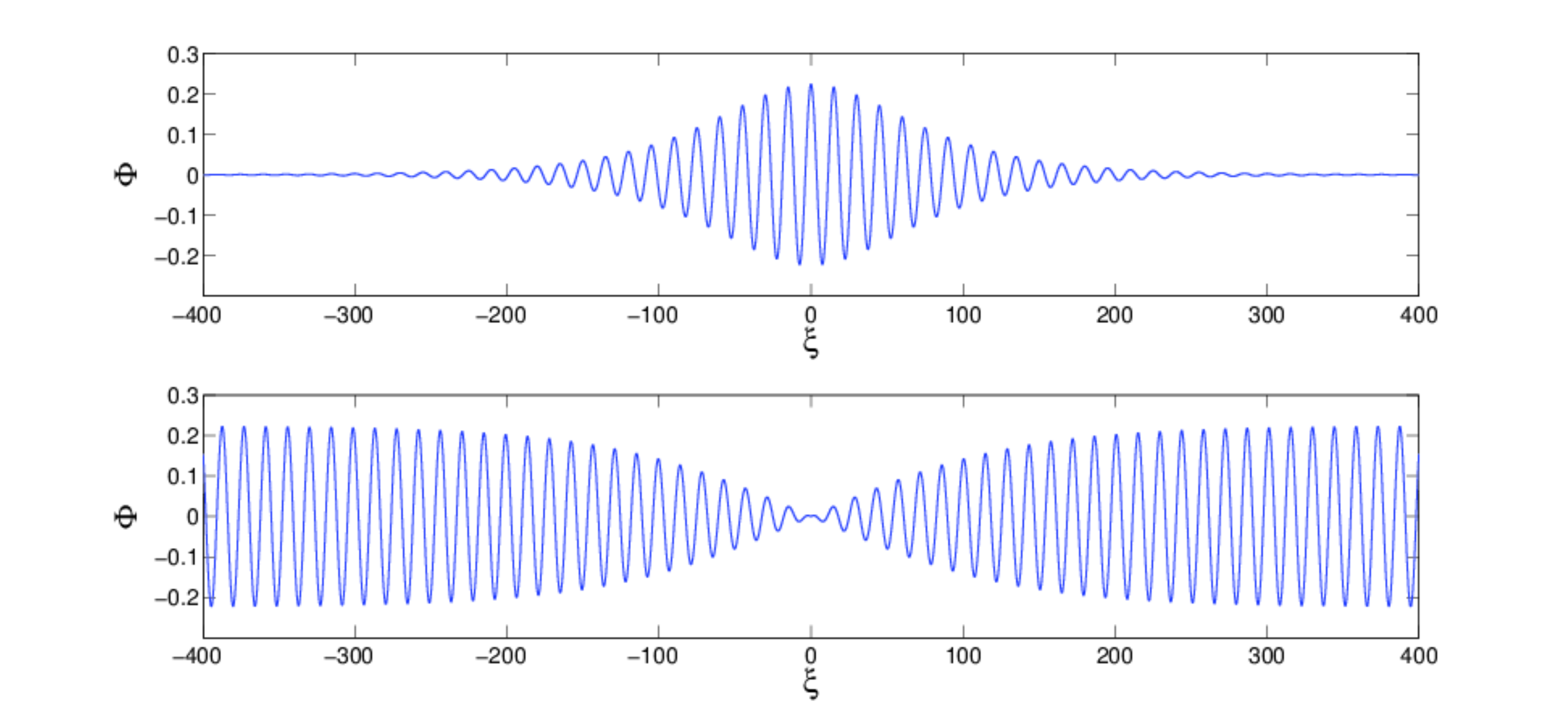}
\caption{The evolution of the bright (at $k=0.5$, upper panel) and dark (at $k=0.3$, lower panel) envelope solitons of Eq. \eqref{NLSE} at $\tau=0$ for $L_n=2,~L_b=6L_n$ and  $\mathcal{H}=0.1$. The other parameter values are $\Psi_{b0}=\Psi_{d0}=0.05,~U=V=0.2$ and $\Omega_0=0.5$.  }
\label{fig:fig-bright-dark}
\end{figure}  

On the other hand, for $P Q < 0$, the  quantum drift-wave packets are modulationally stable which may propagate in the form of  dark-envelope solitons   characterized by a depression of the drift-wave potential around $\xi =0$. Typical form of this solution is given by 
\begin{eqnarray}
&&\Psi=\Psi_{d0}\hskip1pt \text{tanh}^2\left(\frac{\xi-V\tau}{W_d}\right), \notag \\ &&\theta=\frac{1}{2P}\left[V\xi-\left(\frac{V^2}{2}-2PQ\Psi_{d0}\right)\tau\right],\label{dark-envelope}
\end{eqnarray} 
which represents a localized region of hole (void) traveling at a speed $V$. The pulse width $W_d$ depends on the constant amplitude $\Psi_{d0}$ as $W_d=\sqrt{2|P/Q|/\Psi_{d0}}$.
 
\section{Conclusion} We have investigated the amplitude modulation of   one-dimensonal electrostatic drift-wave  packets in a nonunform quantum magnetoplasma in presence of the equilibrium density and magnetic field inhomogeneities, as well as the quantum mechanical (such as tunneling) and the statistical (Fermi-Dirac pressure of degenerate electrons) effects. These effects are shown to modify the dispersion $(P)$ and the nonlinear $(Q)$ coefficients of the nonlinear Schr\"odinger  equation that governs the dynamics of the modulated drift-wave packets. The stable $(PQ<0)$  and unstable $(PQ>0)$ domains of the  the drift-wave number $k$ are obtained for different values of  the quantum parameter $\mathcal{H}$ and the scale lengths of inhomogeneities $L_n$ and $L_b$ for the number density and the magnetic field. It is found that higher values of the quantum parameter $\mathcal{H}$ favor the instability of the drift-wave packets, while the waves may remain stable for a scale length $L_b~(>L_n)$ below a critical value.      It is also seen  that   $P$  is always negative for  $\delta\mathcal{H}^2<1$ and for drift-wave numbers satisfying $k<1$, while $Q$ changes its sign depending on the quantum parameter $\mathcal{H}$ and the scale lengths  $L_n$ and $L_b$ of inhomogeneities. Thus, the formation of the bright    drift-envelope solitons or drift-wave rogons is possible only when $Q < 0$.     On the other hand,  for $Q > 0$, the amplitude modulated drift-wave packet is stable and it propagates in the  form of a dark drift-envelope soliton. { In conclusion,  the drift-wave excitation should be useful in identifying the modulated drift-wave packets that may spontaneously emerge in the core or outer mantles of compact astrophysical objects like white dwarfs, neutron stars that contain the equilibrium number density, electron temperature and magnetic field inhomogeneties at finite scale lengths. The  excitation of bright or dark envelope solitons as well as drift-wave rogons can also be relevant   in  metallic plasmas and should be useful in laser-solid density plasma interaction experiments where the number density of charged particles  varies in the rage $10^{21}-10^{26}$ cm$^{-3}$ with the temperature $T\lesssim10^{7}$ K   and the magnetic field  $B_0\sim10^8$ T or more.}

 \section*{Acknowledgement}
{This research was partially supported by   the SAP-DRS (Phase-II), UGC, New Delhi, through sanction letter No. F.510/4/DRS/2009 (SAP-I) dated 13 Oct., 2009, and by the Visva-Bharati University, Santiniketan-731 235, through Memo No. Aca-R-6.12/921/2011-2012 dated 14 Feb., 2012.}

\bibliographystyle{}

\end{document}